# The Inquisition's Semicolon

Punctuation, Translation, and Science in the 1616 Condemnation of the Copernican System

Christopher M. Graney

Jefferson Community & Technical College

1000 Community College Drive

Louisville, Kentucky 40272

christopher.graney@kctcs.edu

This paper presents high-resolution images of the original document of the 24 February 1616 condemnation of the Copernican system, as being "foolish and absurd in philosophy", by a team of consultants for the Roman Inquisition.  Secondary sources have disagreed as to the punctuation of the document.  The paper includes a brief analysis of the punctuation and the possible effects of that punctuation on meaning.  The original document and its punctuation may also have relevance to public perception of science and to science education.

One of the more notorious statements in the history of science is that one, made on 24 February 1616 by a team of eleven consultants for the Inquisition in Rome, which declared the heliocentric system of Nicolaus Copernicus to be "foolish and absurd in philosophy" and "formally heretical".[1]  Surprisingly, secondary sources disagree regarding the exact wording of this statement, which is written in Latin.  To be more specific, they disagree regarding its exact *phrasing*.  Translations of the statement into English likewise disagree in regards to its phrasing.  The differences in meaning that result are significant.  Thus, this paper includes a copy of the original Inquisition document, with a close translation, and with an analysis of the phrasing of the "foolish and absurd" statement.  Besides being a historical curiosity, the original document and its punctuation may also have some relevance to public perception of science and to science education.

    The consultants' statement was issued as the Inquisition investigated a complaint that had been filed against Galileo in February of 1615.  Galileo had been exonerated, but the Inquisition decided to consult its experts for an opinion on the status of Copernicanism.  However, despite the consultants' statement, the Inquisition issued no formal condemnation[2] of the Copernican system, and the statement was filed away in the Inquisition archives.[3]

---

1  Finocchiaro 1989, 29, 146.

2  However, just after the consultants' report (on 5 March), the Congregation of the Index, the arm of the Vatican in charge of book censorship, issued a decree declaring the Copernican system to be both "false" and "altogether contrary to the Holy Scripture", and censoring books that presented the Copernican system as being more than a hypothesis.

3  Finocchiaro 1989, 28-31, 146-150.

Two decades later, a paraphrase of the statement was made public. This was because, following the trial of Galileo, copies of the 22 June 1633 sentence[4] against him were sent to papal nuncios and to inquisitors around Europe. The sentence, which was written in Italian rather than Latin, noted the opinion of the consultant team and included a paraphrase of their 24 February 1616 statement.[5] Still later, Giovanni Battista Riccioli included a Latin translation of Galileo's sentence in his 1651 *Almagestum Novum*. Riccioli's translation was widely referenced for centuries, but it was a Latin translation of an Italian paraphrase of a Latin original.[6] Modern language translations of the consultants' statement taken from Riccioli simply added a fourth layer of translation. The original statement itself was not published until the middle of the nineteenth century.[7]

However, since then, secondary sources publishing the statement have been inconsistent regarding the punctuation of the original Latin statement (Table 1). Maurice Finocchiaro calls attention to this in his 1989 *The Galileo Affair: A Documentary History*. Finocchiaro includes the following translation of the consultants' statement:

> (1) The sun is the center of the world and completely devoid of local motion.
> Assessment: All said that this proposition is foolish and absurd in philosophy,* and formally heretical since it explicitly contradicts in many places the sense of Holy Scripture, according to the literal

---

4  See Finocchiaro 1989, 287-291.

5  Finocchiaro 2005, 26-42.

6  Finocchiaro 2005, 41.

7  Finocchiaro 2005, 241.

> meaning of the words and according to the common interpretation and understanding of the Holy Fathers and the doctors of theology.
>
> (2) The earth is not the center of the world, nor motionless, but it moves as a whole and also with diurnal motion.
> Assessment: All said that this proposition receives the same judgment in philosophy and that in regard to theological truth it is at least erroneous in faith.[8]

However, at the point noted by the (added) asterisk, Finocchiaro has an endnote that reads as follows:

> At this point, between the word "philosophy" (*philosophia*) and the phrase "and formally heretical" (*et formaliter haereticam*), the original text in the Vatican manuscripts (folio 42r) shows a semicolon; Favaro (19: 321) has a comma; and Pagano (1984, p. 99) has no punctuation. While rules and practices for punctuation were not as developed and strict in the seventeenth century as they are today, and while the Pagano volume is extremely valuable and generally reliable, in this particular instance there seems to be no justification for Pagano's transcription, which conveys the impression that biblical contradiction is being given as a reason for ascribing both philosophical-scientific falsehood and theological heresy.[9]

A Google book search on the exact Latin phrase "absurdam in philosophia et formaliter" (such searches do not distinguish punctuation) will turn up dozens of sources, published in differing languages at differing times, that give the Latin of the consultants' statement. The search results illustrate that different sources indeed give the different punctuations described by Finocchiaro. There is usually either no punctuation or just a comma. Only a handful show a semicolon between *philosophia* and *et formaliter haereticam,* including Gebler 1877,[10] Grisar 1882,[11] and Costanzi 1897.[12]

---

8 Finocchiaro 1989, 146.
9 Finocchiaro 1989, 344 (note 35).

The Pagano work that Finocchiaro cites is a good illustration of the punctuation variation. The 1984 edition has no punctuation between *philosophia* and *et formaliter haereticam*.[13] The 2009 edition has a comma (see Table 1).[14] These works are published out of Vatican City.

A similar search on the phrase "absurd in philosophy and formally" reveals many English translations, again with varied punctuation. A few examples:

> The first proposition was declared unanimously to be foolish and absurd in philosophy and formally heretical inasmuch as it expressly contradicts the doctrine of Holy Scripture in many passages, both in their literal meaning and according to the general interpretation of the Fathers and Doctors.[15]
>
> All said that this proposition is foolish and absurd in philosophy, and formally heretical since it explicitly contradicts in many places the sense of Holy Scripture, according to the literal meaning of the words and according to the common interpretations and understanding of the Holy Fathers and the doctors of theology.[16]

---

[10] Gebler 1877, 47-48.
[11] Grisar 1882, 38.
[12] Costanzi 1897, 221.
[13] Pagano 1984, 99.
[14] Pagano 2009, 42-43.
[15] Langford 1992, 89.

[The proposition] was declared 'foolish and absurd in philosophy, and formally heretical, inasmuch as it expressly contradicts the doctrine of the Holy Scripture in many passages, both in their literal meaning and according to the general interpretation of the Fathers and Doctors.'[17]

All declared the said proposition to be foolish and absurd in philosophy and formally heretical, because it expressly contradicts the doctrine of the Holy Scripture in many passages, both in their literal meaning and according to the general interpretation of the Fathers and the Doctors of the Church.[18]

As Finocchiaro has pointed out, the punctuation creates significant differences in meaning. The question is, what does the original document actually say? Figures 1 and 2 show images of the original document, courtesy of the Vatican Secret Archives. Figure 3 shows a processed image. The resolution of Figures 2 and 3 is sufficient to withstand significant enlargement. These images show that there is punctuation between *philosophia* and *et formaliter haereticam*, and it appears to be a comma with a somewhat elongated dot above it.

That this comma and dot comprise a semicolon is bolstered by the parallel structure of the consultants' statement. Two propositions are being assessed. Both assessments follow the same form "all have

---

said...", followed by a statement regarding Philosophy (capitalized in both instances), and then a statement regarding religion.  The parallel structure found in the overall statement, along with the existence of a semicolon following *Philosophia* in the assessment of the *second* proposition (which numerous sources show, including both Pagano editions[19] -- see Table 1), supports the marks following *Philosophia* in the assessment of the *first* proposition also being a semicolon.[20]

Thus below are both the original Latin[21] of the statement, and our[22] translation of the statement into English -- a translation intended to hew closely to the original in terms of structure, grammar, punctuation, use of cognates, etc.

| | |
|---|---|
| *Sol est centrum mundi, et omnino immobilis motu locali.* | The sun is the center of the world, and entirely immobile insofar as location movement.[23] |
| *Censura: Omnes dixerunt dictam propositionem esse stultam et absurdam in* | Appraisal: All have said the stated proposition to be foolish and absurd in |

---

19 Pagano 1984, 99-100; Pagano 2009, 42-43.

20 Note that the writer of the 24 February 1616 statement often placed the dots of his *i* letters well to the right of the letters themselves.  This is particularly apparent in the first line of Figures 2 and 3.  There the dots from the *i* letters of *mundi* and *locali* give the comma and period in that line the appearance of a semicolon and a colon.  I thank Roger Ceragioli for his comments on this matter.

21 Latin text as in Pagano 2009, 42-43, but with punctuation from Figures 2 and 3.

22 I thank my wife, Christina Graney, for her assistance in this translation.

23 Translational motion, or movement from location to location.  There is no comment here about the rotational motion of the sun.  I thank Roger Ceragioli for his comments on this matter.

| | |
|---|---|
| *Philosophia; et formaliter haereticam, quatenus contradicit expresse sententiis sacrae scripturae in multis locis, secundum proprietatem verborum, et secundum communem expositionem, et sensum, Sanctorum Patrum et Theologorum doctorum.* | Philosophy; and formally heretical, since it expressly contradicts the sense of sacred scripture in many places, according to the quality of the words, and according to the common exposition, and understanding, of the Holy Fathers and the learned Theologians. |
| *Terra non est centrum mundi, nec immobilis, sed secundum se Totam, movetur, etiam motu diurno.* | The earth is not the center of the world, and not immobile, but is moved along Whole itself, and also by diurnal motion. |
| *Censura: Omnes dixerunt, hanc propositionem recipere eandem censuram in Philosophia; et spectando veritatem Theologicam, adminus esse in fide erroneam.* | Appraisal: All have said, this proposition to receive the same appraisal in Philosophy; and regarding Theological truth, at least to be erroneous in faith. |

As Finocchiaro points out, the consultants' appraisal of the two Copernican propositions as "foolish and absurd in philosophy" is a philosophical-scientific appraisal.  They are saying the Copernican system is "scientifically untenable".[24]  But as he notes, punctuation matters.  This sentence --

> All said that this proposition is foolish and absurd in philosophy and formally heretical, since it explicitly contradicts in many places the sense of Holy Scripture....

-- differs in meaning from this one --

> All said that this proposition is foolish and absurd in philosophy, and formally heretical, since it explicitly contradicts in many places the sense of Holy Scripture....

-- which differs from this one --

> All said that this proposition is foolish and absurd in philosophy; and formally heretical, since it explicitly contradicts in many places the sense of Holy Scripture....

But this last version -- which assesses first that the proposition is scientifically untenable, and separately that it is formally heretical since it contradicts Scripture -- makes little sense if we assume the Copernican system had the weight of scientific evidence behind it.  Perhaps this is why the statement has consistently been presented with altered punctuation.  With the original punctuation it does not read in a manner that conforms to what modern readers believe to have been the case.  But recent research has cast more light on the science behind the opposition to the Copernican system.

---

24 Finocchiaro 1989, 29, 344.

Dennis Danielson and I have recently discussed this science in the January 2014 issue of *Scientific American*.[25] Putting things briefly, anti-Copernicans could claim as their own one of the most prominent astronomers of the time, Tycho Brahe. They could cite careful measurements of star diameters which showed that, were the Copernican system correct, stars would be enormous. The sun compared to even an average Copernican star would be like the period at the end of this sentence compared to a grapefruit. By contrast, under a geocentric system,[26] the sizes of celestial bodies would all fall into a consistent range. The moon would be the smallest celestial body, the sun the largest. The stars would be comparable to, but smaller than, the sun. Copernicans could not argue with the data.[27] They resorted to justifying the absurdly large stars in their system by appealing to Divine Majesty and Omnipotence: an infinitely powerful God could easily make such giant stars. This issue was definitely in play in 1616. Several anti-Copernicans had recently cited the star size problem, including Simon Marius[28] and Georg Locher[29] in 1614, and

---

25 Danielson & Graney 2014.

26 Here I mean a geocentric system as modified by Tycho, so that the planets circled the sun while the sun, moon, and stars circled the earth. Such a system was observationally and mathematically identical to the Copernican system insofar as the sun, moon, and planets were concerned. Thus it was fully compatible with telescopic discoveries, such as the phases of Venus.

27 The apparent sizes of stars would turn out to be spurious (an artifact of optics) but that was not discovered for some time. Both visual and early telescopic observations indicated the sizes of stars to be a problem for the Copernican system. One of the first telescopic observers to state that telescopic observations of star sizes supported the Tychonic system was Simon Marius, in his 1614 *Mundus Jovialis*. See Graney 2010.

28 See previous note.

29 Locher, in his 1614 *Disquisitiones Mathematicae*, includes detailed discussions of telescopic discoveries. It features a diagram of how the phases of Venus indicate it to circle the sun, a detailed diagram of the Jovian system with moons passing through the Jovian shadow, and a detailed rendition of the moon as seen through a Francesco Ingoli[30] in 1616 (just weeks prior to the consultants' statement).[31]  From a modern perspective, invoking the power of God to solve a problem with a scientific theory is indeed scientifically untenable.  Thus, in light of what we now know about opposition to the Copernican system, the consultants' assessment (original punctuation intact) makes more sense.

The interesting historical curiosity that is the consultants' statement and its punctuation may be in several ways a boon for public perception of science and science education.  First, the statement is an interesting story.  People are fascinated by the twists and turns involved in the spread of information, correct or otherwise.  Second, the statement undermines the science denial that has gained a certain popularity in recent years.  "Apollo deniers", "9-11 Truthers", "vaccine deniers", and those who assert science to support the universe being 6000 years old all build their claims on the idea that science is a matter of controversies and cover-ups regarding basic truths; that in science, powerful forces suppress inconvenient scientific discoveries.  Many members of the public, as well as many students in introductory science classes, maintain the unfortunate

---

    telescope.  However, Locher places the new discoveries within the system of Tycho.  One of the principle arguments that he cites against the Copernican system is the problem of star sizes.  See Locher 1614, especially page 28: "Argumenti Nucleus", 2nd argument.

30 Ingoli writes, in an early 1616 essay to Galileo, that the great distance of stars in the Copernican system requires that them "to be of such size, as they may surpass or equal the size of the orbit circle of the Earth itself [stellas fixas tantae magnitudinis esse, ut superent aut aequent magnitudinem ipsius circuli deferentis Terram]".  See Favaro 1890-1909, Vol. 5, 406; Graney 2012c, v2, 19-24, 44.  Galileo believed Ingoli to have been influential in the Vatican's 1616 rejection of the Copernican system.  See Finocchiaro 1989, 155.  Finocchiaro argues that Ingoli's essay was very influential (Finocchiaro 2010, 72).

31 For a detailed discussion of the star-size issue, see Graney 2012a (a paper for specialists), or Graney 2012b (a paper for a general academic audience).

view that science is about such controversies.  By undermining the narrative that, at the beginning of the history of modern science, powerful forces conspired to suppress a scientific idea, declaring it to be "foolish and absurd" *only because* it was religiously inconvenient, the consultants' statement should aid in undermining the entire idea of conspiracy and cover-up that is behind the science denial phenomenon.  Third, the statement is interesting to see.  I had expected the document to be a bumptious masterpiece of calligraphy, with an imposing appearance of formality suitable for an Important Proclamation.  But with its very ordinary (and less-than-clear) script and its heavy use of abbreviation, this document -- containing one of the more notorious statements in the history of science -- has less the look of an Important Proclamation than of the hastily scrawled notes from a less-than-important meeting.  Perhaps had the consultants written more neatly, there would be less confusion about the punctuation.

# TABLE 1
## The text of the 24 February 1616 statement by a team of consultants for the Roman Inquisition, as provided by various sources.

**Source: Favaro 1890-1909, Vol. 19, 321.**

Favaro's work is considered a standard reference. There is a comma between the first *philosophia* and *et formaliter,* and a semicolon between the second *philosophia* and *et spectando.*

> XXIV. PROCESSO DI GALILEO.
>
> Prima: Sol est centrum mundi, et omnino immobilis
> Censura: Omnes dixerunt, dictam propositionem ess[e stultam et absurdam]
> in philosophia, et formaliter haereticam, quatenus contra[dicit expresse sententiis]
> Sacrae Scripturae in multis locis secundum proprietatem
> communem expositionem et sensum Sanctorum Patrum et
>
> 2.ª: Terra non est centrum mundi nec immobilis, s[ed secundum se totam]
> movetur, etiam motu diurno.
> Censura: Omnes dixerunt, hanc propositionem recipe[re eandem censuram in]
> philosophia; et spectando veritatem theologicam, ad minu[s esse in fide erroneam.]
>
> Petrus Lombardus, Archiepiscopus Armacanus.
> Fr. Hyacintus Petronius, Sacri Apostolici Palatii

archive.org

**Source: Langford 1992, 89.**

A more recent author who provides the Latin text, citing Favaro.

> sitions seems to be an attempt to translate the Copernican position into some kind of formal philosophical language.
> 
> [23] "Omnes dixerunt, dictam propositionem esse stultam et absurdam in philosophia, et formaliter hereticam, quatenus contradicit expresse sententiis Sacrae Scripturae in multis locis secundum proprietatem verborum et secundum communem expositionem et sensum Sanctorum Patrum et theologorum doctorum."
> 
> [2a] "Omnes dixerunt hanc propositionem recipere eandem censuram in philosophia; et spectando veritatem theologicam, ad minus in Fide erroneam." Opere, XIX, 321.
> 
> [24] Antonius Cordubensis, *Quaestiones Theologicae* (Venice: 1604), I, q.

Google Books.

**Source: L'Epinois 1867, 35.**

There is no punctuation between *philosophia* and *et formaliter,* or between *philosophia* and *et spectando.*

> 24 februarii 1616, coram infrascriptis Patribus theologis. Prima : sol est centrum mundi et omnino immobilis motu locali. Censura : omnes dixerunt dictam propositionem esse stultam et absurdam in philosophia et formaliter hereticam, quatenus contradicit expresse sententiis sacrae Scripturae in multis locis, secundum proprietatem verborum et secundum communem expositionem et sensum SS. Patrum et theologorum doctorum. Secunda : terra non est centrum mundi nec immobilis, sed secundum se totam movetur etiam motu diurno. Censura : omnes dixerunt hanc propositionem recipere eandem censuram in philosophia et spectando veritatem theologicam ad minus esse in fide erroneam. » Suivent les noms des théologiens. (*Ms. du Procès*, fol. 377 v°.)
> 
> [1] « Die Jovis 25 februarii 1616. Ill. D. cardinalis Mellinus notificavit RR. PP. DD. accessori et commissario S. Officii quod relata censura PP. theolo-

Google Books.

**Source: Roberts 1885, 120.**

There is even less punctuation here than in the previous source.

> 120
>
> Censura: Omnes dixerunt dictam propositionem esse stultam et absurdam in philosophia et formaliter hereticam, quatenus contradicit expresse sententiis Sacræ Scripturæ in multis locis secundum proprietatem verborum et secundum communem expositionem et sensum Sanctorum Patrum et Theologorum Doctorum.
>
> Secunda: Terra non est centrum mundi nec immobilis, sed secundum se totam movetur, etiam motu diurno.
>
> Censura: Omnes dixerunt hanc propositionem recipere eandum censuram in philosophia et spectando veritatem theologicam ad minus esse in fide erroneam.
>
> Petrus Lombardus, Archiepiscopus Armacanus.
>
> Fr. Hyacintus Petronius, Sacri Apostolici palatii magister.

Google Books.

---

**Source: Schönert & Vollhardt 2005, 131.**

No punctuation between *philosophia* and *et formaliter*, but a semicolon between *philosophia* and *et spectando*.

> centrum mundi, et omnino immobilis motu locali. Censura: Omnes dixerunt dictam propositionem esse stultam et absurdam in philosophia et formaliter haereticam, quatenus contradicit expresse sententiis Sacrae Scripturae in multis locis secundum proprietatem verborum et secundum communem expositionem et sensum Sanctorum Patrum et theologorum doctorum. Secunda: Terra non est centrum mundi, nec immobilis, sed secundum se totam movetur, etiam motu diurno. Censura: Omnes dixerunt, hanc propositionem recipere eandem censuram in philosophia; et spectando veritatem theologicam, ad minus esse in fide erroneam«. Vgl. auch das Dekret vom 5. 3. 1616, das dann Copernicus' *De revolutionibus orbium coelestium* und Foscarinis *Lettera* von 1615 (Paolo Antonio Foscarini: Lettera del R.P.M. Paolo Antonio Foscarini Carmelitano Sopra l'Opinione de'

Google Books.

---

**Source: Grisar 1882, 38.**

There are semicolons between *philosophia* and *et formaliter*, and between *philosophia* and *et spectando*. Also note the capitalizations.

> Die Censur des ersten Satzes lautete: Dictam propositionem esse stultam et absurdam in Philosophia; et formaliter hereticam, quatenus contradicit expresse sententiis sacre scripture in multis locis. secundum proprietatem verbor. et secundum communem expositionem et sensum Sanctor. Patr. et Theologor. doctor.
>
> Die Censur des zweiten Satzes war: Hanc propositionem recipere eandem censuram in Philosophia; et spectando veritatem Theologicam, ad minus esse in fide erroneam²).

Google Books.

---

**Source: Gebler 1877, 47-48.**

There are semicolons between *philosophia* and *et formaliter*, and between *philosophia* and *et spectando*. Also note the capitalizations, and the indication of line breaks and abbreviations.

> Infrascriptis Patribus Theologis.
>
> Pª.² Sol est centrũ mundi, et omnino im̃obilis motu locali:
> Censura: Omnes dixerunt dictã propositionẽ ẽe stultã et absurdam | in Philosophia; et formaliter hereticã, quatenus contradicit | expresse sententijs sacre scripture in multis locis. secundũ | proprietatẽ verbor, et secundũ communẽ expositionẽ, et | sensũ, Sanct. Patr. et Theologor. doctor.
>
> 2ª. Terra non est centr. mundi, nec im̃obilis, sed secundũ se | totã, movetur, et̃ motu diurno.
>
> Censura. Omnes dixerunt, hanc propositionẽ recipẽ¹ eandẽ censurã in | Philosophia; et spectando veritatẽ Theologicã, ad minus | ẽe in fide erroneã.

Archive.org

| | |
|---|---|
| **Source: Gebler 1879, 77.**<br><br>**Compare to Gebler above. There is a comma between *philosophia* and *et formaliter*, and a semicolon between *philosophia* and *et spectando*.** | [1] Sol est centru mundi, et omnino immobilis motu locali ;<br>Censura : Omnes dixerunt dicta propositionē ēe stultā et absurdam in Philosophia, et formaliter hereticā, quatenus contradicit expresse sententijs sacre scripture in multis locis. Secundū proprietate verbor, et secundū commune expositionē, et sensū. Sanct. Patr. et Theologor doctor.<br>Terra non est centr. mundi, nec immobilis, sed secundu se tota, movetur et moto diurno.<br>Censura : Omnes dixerunt, hanc propositionē recipē eandē censura in Philosophia ; et spectando veritatē Theologicā, at minus ēe in fide erronea. (Vat. MS. folio 377 ro.)<br>[2] Die Iovis. 25th Februarii. 1616.<br><br>Archive.org |
| **Source: Pagano 2009, 42-43.**<br><br>**There is a comma between *philosophia* and *et formaliter*, and a semicolon between *philosophia* and *et spectando*.** | coram intrascriptis Patribus Theologis.<br>Prima: Sol est centrum mundi, et omnino immobilis motu locali.<br>Censura: Omnes dixerunt dictam propositionem esse stultam et absurdam in Philosophia, et formaliter haereticam, quatenus contradicit expresse sententiis sacrae scripturae in multis locis, secundum proprietatem verborum, et secundum communem expositionem, et sensum, Sanctorum Patrum et Theologorum doctorum.<br>2.ª: Terra non est centrum mundi, nec immobilis, sed secundum se Totam, movetur, etiam motu diurno.<br>Censura: Omnes dixerunt, hanc propositionem recipere eandem censuram in Philosophia; et spectando veritatem Theologicam, adminus esse in fide erroneam.<br>Petrus Lombardus, Archiepiscopus, Armacanus [132] |

Propositiones Censurandæ

Censura facta in Sᵗᵒ Officio Urbis die Mercurij 24 Februarij 1616 coram Infrascriptis Patribus Theologis.

1ª: Sol est centrum mundi, et omnino immobilis motu locali.

Censura: Omnes dixerunt dictam propositionem esse stultam et absurdam in Philosophia; et formaliter hæreticam, quatenus contradicit expresse sententiis sacræ scripturæ in multis locis, secundum proprietatem verborum, et secundum communem expositionem, et sensum Sanctorum Patrum et Theologorum doctorum.

2ª: Terra non est centrum mundi, nec immobilis, sed secundum se totam movetur, etiam motu diurno.

Censura: Omnes dixerunt, hanc propositionem recipere eandem censuram in Philosophia; et spectando veritatem Theologicam, ad minus esse in Fide erroneam.

Petrus Lombardus Archiepiscopus Armacanus

Fr. Hyacinthus Petronius sac. Apost. Pal. Mag.

Fr. Raphael Riphoz theol. magr et Vicar. Generalis ordinis prædicatorum

Fr. Michael Angelus seg.ᵗᵃʳⁱᵘˢ Theologus et consult. Sᵗⁱ Off.

Fr. Hieronymus de Casalimaiori consultor Sᵗⁱ Officij

Fr. Thomas de Lemos

Fr. Gregorius Nunnius Coronel

Bened.ᵗᵘˢ Justinus Societatis Jesu

D. Raphael Rastellius Cler. Reg. doctor Theologus

D. Michael à Neapoli ex congregatione Cassinensi

Fr. Iacobus Tintus socius R.mi P.ris Commissarij Sᵗⁱ Off.

42  35

FIGURE 2: Detail from Figure 1.
This image will withstand significant enlargement using a PDF reader.
© 2014 Archivio Segreto Vaticano, Misc. Arm. X, 204, ff. 42r

FIGURE 2: Processed detail from Figure 1.
This image will withstand significant enlargement using a PDF reader.
© 2014 Archivio Segreto Vaticano, Misc. Arm. X, 204, ff. 42r